\documentclass[11pt]{article}
\usepackage{amssymb}
\usepackage{epsfig}
\begin{document}
\title{{\bf{\Large Entropy current in two dimensional
 anomalous hydrodynamics and a bound on the sum of the response parameters}}}
\author{  {\bf {\normalsize Rabin Banerjee$^1$}$
$\thanks{E-mail: rabin@bose.res.in}},$~$ 
{\bf {\normalsize Shirsendu Dey$^{1,2}$}$
$\thanks{E-mail: shirsendu12@bose.res.in}}
and
 {\bf {\normalsize Bibhas Ranjan Majhi$^{3,4}$}$
$\thanks{E-mail: bibhas.majhi@iitg.ernet.in; bibhas.majhi@mail.huji.ac.il}}$~$
\\
\\{\normalsize $^1$S. N. Bose National Centre for Basic Sciences,}
\\{\normalsize JD Block, Sector III, Salt Lake, Kolkata-700098, India}
\\
{\normalsize $^2$Department of Physics, Kalyani Government Engineering College,}
\\{\normalsize Kalyani, Nadia-74123, India}
\\
{\normalsize $^3$Department of Physics, Indian Institute of Technology Guwahati,}
\\{\normalsize  Guwahati 781039, Assam, India}
\\
{\normalsize $^4$Racah Institute of Physics, Hebrew University of Jerusalem,}
\\{\normalsize Givat Ram, Jerusalem 91904, Israel}
\\[0.3cm]
} 
\maketitle
\begin{abstract}
The exact expression for the entropy current of a fluid in presence of two dimensional gravitational anomalies is given. To make it compatible with the second law of thermodynamics; i.e. positivity of the entropy production rate of a system (which is considered to be fundamental), we find a bound on the sum of the two response parameters ($\bar{C}_1$ and $\bar{C}_2$), in terms of the trace and diffeomorphism anomaly coefficients ($c_w$ and $c_g$). The precise expression, we obtain here, is $\bar{C}_1 + \bar{C}_2\leq 4\pi^2c_w + 8\pi^2 c_g$. Interestingly, when the bound is saturated corresponding to a reversible process, the result reproduces earlier findings obtained by either studying field theory on the cone or using the {\it Israel-Hartle-Hawking} boundary condition. Finally, possible physical implications and connections with the existing approaches are addressed.     
\end{abstract}
\section{Introduction}
Relativistic fluid dynamics is the effective description of microscopic field theory, valid over spatio-temporal length scales large compared to the mean free path of any interacting system \cite{Landau}. It turns out that the relativistic fluid is characterized by several macroscopic fluid parameters, like the comoving velocity and the thermodynamic variables -- energy density, pressure, temperature etc. In the case of the usual fluids, the dynamical equations are simply represented by the conservation equations for the stress tensor and the currents (for gauge theory). The basic phenomenological inputs are the constitutive relations where stress tensor and the currents, which satisfy the required conserved equations, are written in terms of the above mentioned fluid variables. 

Incidentally, these constitutive relations must be compatible with all the consistency conditions that arise from the degrees of freedom of the system. In most of the cases, entropy plays an important role for the thermodynamic description. Usually one defines a current, known as {\it entropy current}, to represent the entropy of the system \cite{Landau}. For example, in the hydrodynamic regime, one can visualise the fluid--gravity correspondence as an extension of black hole thermodynamics where the horizon entropy can be represented by a local entropy current \cite{Hubeny:2011hd}. For any such thermodynamic description, the second law of thermodynamics plays an important role. More precisely, the entropy current of any system is determined subject to the condition that it satisfies all the fundamental laws of thermodynamics. Usually in hydrodynamics, one uses a local version of second law of thermodynamics to obtain it. This also ensures compatibility with the constitutive equations \cite{Hubeny:2011hd}.
A word about our definition of the entropy current is in order. It should be pointed out that in usual hydrodynamical framework the entropy current is constructed by the gradient expansion approach under the condition that its divergence is non negative. However, such an analysis does not define the entropy current uniquely beyond the leading order \cite{Romatschke:2009kr}. Thus, within the standard hydrodynamical framework, the present definition seems to be the only possiblilty. Of course there are other ways to define it using holgraphic principles. This is basically based on the bulk-boundary map. Here the current is defined in terms of the area of the horizon of a black brane which is treated as the bulk portion of the theory. The interesting fact of this method is that the positivity of the divergence of the current is automatically satisfied by the usual area increase theorem. Moreover, one can also extend such a concept to other horizons like "apparent horizons" \cite{Booth:2010kr}.

  Recently, people explored the role played by both the gravitational as well as the gauge anomalies in hydrodynamics. It turned out that it has important consequences in the fluid--gravity correspondence. This led to diverse approaches for discussing anomalous hydrodynamics over the last few years \cite{Son:2009tf}--\cite{Megias:2014mba}. In the presence of anomalies, the dynamics changes leading to modifications in the constitutive equations with higher order derivatives in fluid variables. The entropy current is obtained by adopting the derivative expansion method. The various coefficients in this expression are fixed by demanding consistency with the local form of the second law of thermodynamics. However it is useful to mention that these are all approximate results which are valid upto some order in derivatives of fluid variables (For details and a list of references, see \cite{Hubeny:2011hd}).

  Very recently, we have shown that the treatment of the $(1+1)$ dimensional case is rather different from others. The effective action itself can be exactly found because the two-dimensional metric, in general, can be expressed in a conformally flat form. From a knowledge of the effective action the gauge current or the stress tensor may be computed by taking appropriate functional derivatives. In this way recourse to derivative expansion is bypassed and one may obtain results in an exact form. In particular the two dimensional anomalous constitutive relations and properties related to it were found by us in a set of papers \cite{Banerjee:2013qha,Banerjee:2013fqa,Bibhas:2014rp,Dey:2014,Bibhas:2014vc}.
  
  In the present paper we follow a similar strategy and obtain the entropy current, in the presence of pure gravitational anomaly, without adopting the derivative expansion method. Naturally, this result is connected with the structure of the constitutive relation involving the stress tensor, in the presence of the gravitational anomaly. Since the constitutive relation involves both the response parameters $(\bar{C}_1,\bar{C}_2)$ as well as the coefficients of the conformal $(c_w)$ and gravitational $(c_g)$ anomalies,\footnote{The conformal anomaly is related to the non-vanishing trace of the stress tensor while the gravitational anomaly emerges from the non-vanishing of the divergence of the stress tensor \cite{Bardeen:1984pm}-\cite{Gaume-witten:1984}.}  the entropy current also involves them. By requiring compatibility with the second law of thermodynamics we obtain a bound on the sum of the response parameters in terms of the anomaly coefficients. When this bound is saturated, corresponding to a reversible process, the result agrees with the individual values of the parameters $(\bar{C}_1,\bar{C}_2)$ obtained earlier by either studying field theory on a cone \cite{Jensen:2012kj} or adopting the {\it{Israel-Hartle-Hawking}} vacuum as a boundary condition \cite{Bibhas:2014rp,Dey:2014}.

The paper is organized in the following way. In section  \ref{sec2}, the anomalous Ward identities and required constitutive relations for energy momentum tensor are given. Let us mention that we do not present the calculations explicitly to find the anomalous constitutive relation, rather take advantage of the results obtained earlier in \cite{Banerjee:2013qha,Banerjee:2013fqa,Bibhas:2014rp}. Then in the next section we elaborately calculate the entropy current and its divergence. Using the second law of thermodynamics, section \ref{sec4} logically gives the cherished relation between the response parameters and the anomaly coefficients. Finally we conclude in section \ref{sec5}.
\section{\label{sec2}Setup: anomalous constitutive relations}
Hydrodynamics in the presence of anomaly has been discussed on many occasions for the last few years \cite{Son:2009tf}--\cite{Megias:2014mba}. Particularly in $(1+1)$ dimensions this has received considerable attention \cite{Dubovsky:2011sk,Jains:2013}, \cite{Jensen:2012kj}--\cite{Banerjee:2013fqa},  \cite{Jensen:2013rga}, \cite{Bibhas:2014rp}--\cite{Chang:2014jna}. We shall not give any detailed calculation in this section, rather will refer relevant references and give the appropriate results. It is well known that, at the quantum level, all symmetries present classically may not be preserved. In general the covariant diffeomorphism and conformal anomaly for energy momentum tensor is given by \cite{Bardeen:1984pm}--\cite{Bertlmann:1996xk}:
\begin{equation}
\nabla_b T^{ab} =  c_g\bar{\epsilon}^{ab}\nabla_b R; \,\,\,\ T^{a}_{a} = c_w{R}~,
\label{diffchiral}
\end{equation}
where $R$ is the two dimensional Ricci scalar, $c_g$ and $c_w$ are diffeomorphism and conformal anomaly coefficients respectively. It is simple to motivate this result. Arguments based on dimensionality and covariance immediately yield the functional structures of the diffeomorphism or conformal anomaly. The normalisation constants may be fixed for a specific problem by doing a loop calculation or using topological methods \cite{Bertlmann:1996xk}. 

  The most general static background metric for hydrodynamics in $(1+1)$ dimensions can be taken as:
\begin{equation}
ds^2 = - e^{2\sigma(r)}dt^2 + g_{11}(r) dr^2~,
\label{effectivemetric}
\end{equation}
It must be noted that one would have a cross term in the above. As the present case is static, it is always possible to impose a coordinate transformation which leads to vanishing of the cross term so that the metric is diagonal. This does not violate the number of degrees of freedom for the metric coefficients. Because, in general, in two dimensions the number of independent metric coefficients is three. Now since there is one coordinate transformation, we shall have ($3-1)=2$ independent metric coefficients which is precisely reflected in (\ref{effectivemetric}).

   To find the constitutive relation; i.e the energy-momentum tensor, which leads to the anomaly expressions (\ref{diffchiral}), it is necessary to solve them under the background (\ref{effectivemetric}). Next using the relations between the fluid variables (like temperature, velocity, chemical potential, etc.) and the metric coefficients in the comoving frame, one writes the constitutive relation for the fluid. Interestingly, this program is very useful in ($1+1$) dimensional theory as the anomaly equations (\ref{diffchiral}) are exactly solvable and hence we can obtain the exact expression of anomalous stress tensor. Indeed the two dimensional effective action itself can be exactly computed. In this respect, the procedure is much more useful and economical than the approximate derivative expansion approach. We have already shown all such features in our earlier works \cite{Banerjee:2013qha,Banerjee:2013fqa,Bibhas:2014rp,Dey:2014}. So, instead of going into the details, let us just write the general form of the constitutive relation. The stress-tensor is given by 
 \cite{Bibhas:2014rp}: 
\begin{eqnarray}
\label{chi2}
T_{ab} &=& \left[2c_w\left(u^c \nabla^d - u^d\nabla^c\right)\nabla_c u_d + 2\bar{C}_1T^2 \right]{u}_a{u}_b
\nonumber
\\
&-& \left[2c_g\left(u^c \nabla^d - u^d\nabla^c\right)\nabla_c u_d + \bar{C}_2T^2 \right]({u}_a\tilde{u}_b+\tilde{u}_au_b)
\nonumber
\\
&+& \left[\bar{C}_1T^2-c_w\left(u^c \nabla^d \nabla_d u_c \right)\right]g_{ab}~.
\end{eqnarray}
The energy momentum tensor (\ref{chi2}), which contains the higher order derivative terms, is responsible for the anomalies (\ref{diffchiral}). $u^a$ is the comoving fluid velocity which satisfies the timelike condition $g_{ab}u^au^b = -1$ and $\tilde{u}_a = \bar{\epsilon}_{ab}u^b$ is the dual of it. $T$ is the temperature of the fluid which is related to the equilibrium temperature $T_0$ by a Tolman like relation $T=e^{-\sigma}T_0$. $\bar{C}_1$ and $\bar{C}_2$, known as response parameters of the fluid, are related to the integration constants appearing in the solutions of anomaly equations (\ref{diffchiral}).
 For a detailed analysis and the meaning of the symbols, see \cite{Banerjee:2013fqa,Bibhas:2014rp}. Note that in (\ref{chi2}) we have zeroth order and then second order derivative terms. Whereas there is no first order term. Actually the first order terms are responsible for the viscous part of the fluid, i.e the shear viscosity ($\zeta$) and the bulk viscosity ($\eta$). For the appearance of such terms, the minimum space-time dimension of the theory should be three. But since our theory is described in ($1+1$) dimensions, bulk and shear viscosity coefficients vanish ($\zeta = \eta=0$). Moreover, remember that the expression (\ref{chi2}) is exact; i.e. we do not have any other higher order corrections. Also, it may be worthwhile to mention an important feature of (\ref{chi2}) that the response parameters appear in zeroth order  terms whereas the anomaly coefficients appear in second order terms.

  Within this setup, the next interesting question is: What is the expression for the entropy current? This is important as entropy plays a significant role in the thermodynamic paradigm of fluid. Since the stress tensor (\ref{chi2}) is an exact one, it is in principle possible to find the entropy current which is also exact. This will be precisely done below. Moreover, validity of the second law of thermodynamics implies that the divergence of entropy current should be greater than (for an irreversible system) or at least equal to zero (for a reversible system). If one imposes such a fundamental constraint, we shall see that the sum of the two response parameters, $\bar{C}_1$ and $\bar{C}_2$, must satisfy a bound in terms of the two anomaly coefficients, $c_w$ and $c_g$.

 Before going into the main computation, we shall end this section by giving the following expressions, valid for the metric (\ref{effectivemetric}), which will be needed for the next two sections: 
\begin{equation}
u^{a} = \Big(e^{-\sigma},0\Big); \,\,\ \tilde{u}^{a} = \Big(0,\frac{1}{\sqrt{g_{11}}}\Big)~.
\label{u}
\end{equation}  
Remember that these are evaluated in ($t,r$) coordinate system.

\section{\label{sec3}Entropy current and its divergence }
 Consider a basic thermodynamic system with extensive quantities like entropy ($S$), volume ($V$), energy ($E$) and intensive ones like temperature ($T$), pressure ($p$) and velocity ($u$). Correspondingly, one can define the densities; i.e. the quantity per unit volume, for energy and entropy, as $\epsilon$ and $s$.
 In case of a covariant theory for relativistic hydrodynamics, quantities like $(T,p,V)$ are associated to scalars, which are measured by an observer at rest with respect to the fluid. Other variables $s,u$ are associated to vectors like $s^a$ and $u^a$, where $u^a$ is the comoving velocity of the fluid as defined earlier and $s^a$ is the entropy current.
 The entropy current for an ideal fluid, in general, can be written as:
\begin{eqnarray}
\label{entc}
s^a_{ideal}=s_{(ideal)}u^a~,
\end{eqnarray}
where $s_{(ideal)}$ is the corresponding entropy density. For completeness, let us mention that $s_{(ideal)}$ is given by
\begin{eqnarray}
\label{ideal}
s_{(ideal)}=(\epsilon + p)/T.
\end{eqnarray}
 This structure of the ideal entropy current can be explained as follows. For the ideal fluid the stress tensor $T_{(ideal)ab} =(\epsilon+p)u_au_b+pg_{ab}$, energy current can be defined as $\epsilon^a =-T^{ab}_{(ideal)}u_b$. Then the entropy current in this case can be defined as $T s^a_{(ideal)} = \epsilon^a+pu^a$. Now using $g_{ab}u^au^b=-1$ one can find that $\epsilon^a = \epsilon u^a$ and hence we have (\ref{entc}) 
where $s_{(ideal)}$ is defined in (\ref{ideal}).

 Entropy current for the ideal fluid is a conserved quantity. For instance, since the background metric is static, we have $\nabla_a (s_{(ideal)}u^a) = 0$. In presence of anomalus terms in the fluid action, the current must be constructed such that it is compatible with the second law of thermodynamics.  
  In the literature there exist some formalisms to find the higher order corrections, namely the dissipative terms, in the entropy current. Since $T^{ab}$ contains second order terms (see Eq. (\ref{chi2})), it may be possible to adopt these formalisms to find the corresponding entropy current. But this is not possible. The reasons are as follows. First of all, here we cannot use Landau-Lifshitz \cite{Landau} or Eckart's theory \cite{Eckart:1940te} as they are first order formalisms. Whereas, in our case the stress-tensor contains second order terms. Although there is a second order formalism by Israel and Stewart \cite{Israel}, but that is restricted to the cases where one has $\nabla_a T^{ab} = 0$ and $\tau_{ab} u^a = 0$, with $\tau^{ab}$ as the higher order corrections or the dissipative contribution. Unfortunately, none of the conditions are valid for the present case due to the presence of the anomalies with the identification $\tau_{ab}\equiv T_{ab}$. The other option is to use the derivative expansion approach. Since it will not give an exact result, we do not want to adopt it. Therefore, in the absence of any appropriate formalism to find the entropy current for our second order case, the entropy current for anomalous fluid will be determined following the prescription of the ideal fluid, as described above.
 We define entropy current as {\footnote{Note the crucial distinction from the ideal case (\ref{entc}) which involves only
the velocity $u$ but not its dual $\tilde u$. As subsequently shown the
nontrivial contribution comes only from the dual part. Thus the ideal
sector does not contribute.}}
 \begin{eqnarray}
\label{TSa}
 Ts^a = -\tau^{ab}(u_b+\tilde{u}_b)= -T^{ab}(u_b+\tilde{u}_b).
\end{eqnarray}
The stress tensor derived from anomlaous effective action is contained in $\tau^{ab}$ which has to be contracted with both the velocity and its dual.
This gives an expression for the entropy current for anomalous hydrodynamics in terms of (\ref{chi2}).
As we have mentioned earlier, since we cannot use Eckart's theory or Israel-Stewart theory, the above entropy current has been given following the ideal fluid prescription. In this sense (\ref{TSa}) is not the only possible choice. However, in the absense of any definite prescription we take (\ref{TSa}) as the expression of entropy current for the present analysis. Interestingly, we shall notice in our subsequent analysis that such a definition is consistent with existing results.

    After obtaining the entropy current, it is necessary to calculate its divergence to make it consistent with the second law of thermodynamics, because the law implies that for the total manifold one should have $\nabla_a s^a\geq 0$. The inequality sign comes only for an irreversible process while the equality holds for the reversible one. 
 Now using (\ref{TSa}) we obtain
  \begin{eqnarray}
  \label{eckart}
  \nabla_a s^a= -\nabla_a\left(\frac{\tau^{ab}(u_b+\tilde{u}_b)}{T}\right)~.
\end{eqnarray} 
Any quantity, like $\nabla_a(\Phi u^a)$ where $\Phi$ is an arbitrary time independent scalar, vanishes for the background (\ref{effectivemetric}) because $u^a$ has only a time component (see, Eq. (\ref{u})) and so 
\begin{equation}
\nabla_a(\Phi u^a) = (1/{\sqrt{-g}})\partial_t({\sqrt{-g}}\Phi u^{t})
\end{equation}
which is equal to zero since everything is time independent. Similarly, one concludes 
\begin{equation}
u^a\nabla_a \Phi =0~.
\end{equation}
These have been used in the above and will be used repeatedly in the subsequent calculations.

   For the convenience of computations, let us rewrite (\ref{chi2}) in the following form:
\begin{eqnarray}
&& T^{ab} = \mathcal{A}{u}^a{u}^b + \mathcal{B}({u}^a\tilde{u}^b+\tilde{u}^au^b) + \mathcal{C}g^{ab}
\label{T}
\end{eqnarray}
where
\begin{eqnarray}
&&\mathcal{A}=2c_w\left(u^c \nabla^d - u^d\nabla^c\right)\nabla_c u_d+ 2\bar{C}_1T^2 
\nonumber
\\
&&\mathcal{B}=- \left[2c_g\left(u^c \nabla^d - u^d\nabla^c\right)\nabla_c u_d + \bar{C}_2T^2 \right]
\nonumber
\\
&&\mathcal{C} = \bar{C}_1T^2 -c_w\left(u^c \nabla^d \nabla_d u_c \right)~.
\label{ABC}
\end{eqnarray}
Therefore, use of $u^au_a=-1$, $\tilde{u}^au_a =0$ and $\tilde{u}^a\tilde{u}_a=1$ leads to  
\begin{eqnarray}
Ts^a = (\mathcal{A}-\mathcal{B}-\mathcal{C})u^a + (\mathcal{B}-\mathcal{C})\tilde{u}^a.
\end{eqnarray}
 Substitution in (\ref{eckart}) yields,
\begin{eqnarray}
\nabla_as^a = \nabla_a\Big(\frac{\mathcal{B}-\mathcal{C}}{T}\tilde{u}^a\Big)
\label{divergence}
\end{eqnarray}
since the term proportional to $u^a$ drops out for reasons discussed below (\ref{eckart}).

This is the final expression for the divergence of the entropy current in two dimensions in presence of gravitational anomalies. In the next section, we shall impose the second law of thermodynamics to obtain a relation between the response parameters and the anomaly coefficients. Note that in the final result $\bar{C}_1$ and $c_w$ appear through $\mathcal{C}$ where as $\bar{C}_2$ and $c_g$ appear through $\mathcal{B}$. Hence we expect a condition among these parameters which will ensure compatibility with the second law of thermodynamics.

\section{\label{sec4}Second law of thermodynamics and a bound on the sum of the response parameters}
     In the above section, an expression for the entropy current was given in presence of the anomalies. This current has to be consistent with the fundamental laws of physics. Such a criteria imposes constraints on the parameters appearing in the expression (\ref{TSa}). Here we shall impose the condition that our current is consistent with the second law of thermodynamics. For that one needs to calculate the divergence of the current. This has been evaluated earlier and will be the starting point of the present analysis.

  Note that $\nabla_as^a$ is defined for the whole manifold $\mathcal{M}$. Therefore the second law of thermodynamics (increase of the rate of the entropy current); i.e. $\nabla_as^a\geq 0$ can be cast in the following form:    
 \begin{eqnarray}
 \label{Gdiv}
 \int_\mathcal{M} (\nabla_a{s^a})\sqrt{-g} d^2x\geq 0~,
 \end{eqnarray}
by integrating over the full manifold. Here $\mathcal{M}$ is two dimensional. Now use of the value of the divergence of the entropy current, given by (\ref{divergence}), yields
\begin{equation}
\int_\mathcal{M}  \sqrt{-g} d^2x \nabla_a\Big(\frac{\mathcal{B}-\mathcal{C}}{T}\tilde{u}^a\Big)\geq 0~.
\label{M}
\end{equation}
Using Gauss's theorem, we convert the above into a line integral:
\begin{eqnarray}
\label{Gauss1}
 \int_\mathcal{M}  \sqrt{-g} d^2x \nabla_a\Big(\frac{\mathcal{B}-\mathcal{C}}{T}\tilde{u}^a\Big)= \oint_\Sigma \frac{\mathcal{B}-\mathcal{C}}{T}\tilde{u}^an_a \sqrt{|h|}dx \geq 0,
\end{eqnarray}
where $\Sigma$ is the closed boundary enclosing $\mathcal{M}$. $n^a$ is the unit normal on $\Sigma$ and $h$ is the determinant of the induced metric of the boundary. For the present case, the boundary $\Sigma$ can be divided into four parts: two spacelike surfaces ($t=constant$) and two timelike surfaces ($r=constant$). We denote the spacelike surfaces as $\Sigma_{t_1}$ and $\Sigma_{t_2}$ whereas the timelike surfaces as $\Sigma_{r_1}$ and $\Sigma_{r_2}$. Then the unit normals of $\Sigma_t$ surfaces will satisfy $n^an_a=-1$ (timelike condition) while for others, it will be $n_an^a=+1$ (spacelike condition). For the present case, the metric of $\mathcal{M}$ is given by (\ref{effectivemetric}) which has a Killing horizon $r=r_0$ (say). Therefore, for the accessible region of spacetime one $r=constant$ surface can be taken as the horizon ($\mathcal{H}$) and other can be considered to be at infinity. Keeping this in mind we denote $\Sigma_{r_1}\equiv \mathcal{H}$ and $\Sigma_{r_2} \equiv \Sigma_{\infty}$. Then the right hand side is decomposed into four terms as below: 
 \begin{eqnarray}
 \label{Gsur}
 \oint_\Sigma \frac{\mathcal{B}-\mathcal{C}}{T}\tilde{u}^an_a \sqrt{|h|}dx &=& \int_{\Sigma_{t_1}}  \frac{\mathcal{B}-\mathcal{C}}{T}\tilde{u}^an_a\sqrt{|h|}dr-\int_{\Sigma_{t_2}}\frac{\mathcal{B}-\mathcal{C}}{T}\tilde{u}^an_a\sqrt{|h|}dr
\nonumber
\\
&+& \int_{\mathcal{H}}  \frac{\mathcal{B}-\mathcal{C}}{T}\tilde{u}^an_a\sqrt{|h|}dt-\int_{\Sigma_{\infty}}
  \frac{\mathcal{B}-\mathcal{C}}{T}\tilde{u}^an_a\sqrt{|h|}dt~.
 \end{eqnarray}
The relative negative sign appears in between two similar consecutive terms  as the normals are taken to be opposite in direction. Now note that, for the first two integrals the unit normals have only time components (as $n^a$ is proportional to $u^a$). On the other hand, as we can see from (\ref{u}), the non-zero component of $\tilde{u}^a$ has only the space part. Therefore, the first two terms in the right hand side of (\ref{Gsur}) will vanish. Moreover, as one of the spatial surface is at infinity, the contribution can be set to zero with  the standard assumption that the fields are vanishing at asymptotic infinity. Under these circumstances, (\ref{Gsur}) reduces to{\footnote{In the case of usual spacetimes (i.e. without horizon), the timelike boundary is always at infinity. So there will not be any contribution from it. The actual contribution comes from the space integration; i.e. from the first two terms of (\ref{Gsur}). But the space-time with horizon is little different. Since one can take one timelike boundary as the horizon, which is at finite distance, we have a non-zero value from it. This has precisely occurred here.}}
 \begin{eqnarray}
  \label{enth}
  \oint_\Sigma \frac{\mathcal{B}-\mathcal{C}}{T}\tilde{u}^an_a \sqrt{|h|}dx=\int_{\mathcal{H}}\frac{\mathcal{B}-\mathcal{C}}{T}\tilde{u}^an_a\sqrt{|h|}dt.
  \end{eqnarray}
Hence (\ref{Gauss1}) implies that
\begin{equation}
\int_{\mathcal{H}}\frac{\mathcal{B}-\mathcal{C}}{T}\tilde{u}^an_a\sqrt{|h|}dt\geq 0~.
\label{sigma}
\end{equation} 
The above integration can be evaluated as the integral is time independent. 

     Before proceeding further, let us comment on (\ref{sigma}). Note that (\ref{sigma}) is a consequence of (\ref{divergence}). In the usual case the entropy current $s^a$ is proportional to the velocity field $u^a$. Since for a time independent background, $u^a$ has only time component (see (\ref{u})), one must have vanishing divergence. But in our present case $s^a$ contains both $u^a$ and its dual $\tilde{u}^a$ (see (\ref{TSa})). Such a structure appears due to the presence of anomalies. Note for instance, that the anomalous stress tensor (\ref{chi2}) contains both these fields. For the same background $\tilde{u}^a$ has only radial component (see (\ref{u})). Thats why we do not have vanishing divergence, which is precisely reflected in (\ref{divergence}). Thus, although our background is time independent, still we have a non-zero divergence. Thus integrated version of the local law of thermodynamics leads to (\ref{sigma}).

Considering the limit of integration of time from zero to the periodicity of the Euclidean time, which is the inverse of equilibrium temperature $T_0$, we obtain
\begin{equation}
\lim_{r\rightarrow r_0}\frac{\left(\mathcal{B}-\mathcal{C}\right)T_0^{-1}}{T}\tilde{u}^an_a\sqrt{|h|}\geq 0~.
\label{bound}
\end{equation}
For our particular choice of background metric (\ref{effectivemetric}), one can see that $\tilde{u^a}=(0,\frac{1}{\sqrt{g_{11}}})$ and $n_a=(0,\sqrt{g_{11}})$ while $\sqrt{|h|}=e^{\sigma}$. Therefore, (\ref{bound}) reduces to
    \begin{eqnarray}
    \label{euclT}
    \lim_{r\rightarrow r_0}\frac{\mathcal{B}-\mathcal{C}}{T^2}\geq 0,
    \end{eqnarray}
where $T=T_0e^{-\sigma}$ has been used.

  Now, as $\mathcal{B}$ and $\mathcal{C}$ are given by (\ref{ABC}), for the metric (\ref{effectivemetric}) they can be expressed as
\begin{eqnarray}
\mathcal{B} &=& -\frac{c_g}{g_{11}^2} \Big[2\sigma''g_{11}-\sigma'g_{11}'\Big]-\bar{C}_2T^2~;
\nonumber
\\
\mathcal{C} &=& \bar{C}_1T^2 - c_w \frac{\sigma'^2}{g_{11}}~. 
\label{B}
\end{eqnarray}
Substitution of these in (\ref{euclT}) implies,
\begin{eqnarray}
\label{divh}
&&\lim_{r\rightarrow r_0}\Big[\Big\{-\frac{c_g}{g_{11}^2T^2} \Big(2\sigma''g_{11}-\sigma'g_{11}'\Big)-\bar{C}_2\Big\} - \Big\{\bar{C}_1 - c_w \frac{\sigma'^2}{g_{11}T^2}\Big\} \Big]\geq 0
\nonumber
\\
&&\Rightarrow \bar{C}_1+\bar{C}_2\leq \lim_{r\rightarrow r_0} \Big[c_w \frac{\sigma'^2}{g_{11}T^2} - \frac{c_g}{g_{11}^2T^2} \Big(2\sigma''g_{11}-\sigma'g_{11}'\Big) \Big]~.
\end{eqnarray}
Till now our analysis depends only on the background metric (\ref{effectivemetric}) with no other extra information. Now to calculate the right hand side of (\ref{divh}) explicitly we shall assume the metric to be a solution of Einstein's equations. Thus this is a reasonable assumption follows from the following observations. We know for a static background the Killing horizon ($r_K$) is defined by $e^{2\sigma}|_{r_K}=0$. This information is not sufficient to evaluate the right hand side of (\ref{divh}), because it contains the other metric coefficient $g_{11}$. To handle this term we need extra information. It is well known that, if we consider (\ref{effectivemetric}) as a black hole solution of Einstien's equations, then it must have an event horizon $(r_E)$ which is determined by the equation $1/g_{11}|_{r_E}=0$. In this case the two horizons coincide \cite{Carter}. We denote this horizon as $r_K=r_E=r_0$. The rest of the analysis will be based on this assumption. Furthermore, let us make a note that the background metric, in principle, can be taken as a solution of Einstien's equations with or without the stress tensor (\ref{chi2}) as a source on the right hand side. The difference between these two cases appears in the explicit expressions of the metric coefficients. In this analysis we do not need such information and hence one does not worry about the explicit form of the solutions.

   To proceed, let us first write the right hand side of (\ref{divh}) in terms of metric coefficients. For simplicity of notation, define  $e^{2\sigma}\equiv f(r)$ and $\frac{1}{g_{11}}\equiv g(r)$. Then explicit calculation yields
\begin{eqnarray}
\label{barc23}
&&c_w\frac{{\sigma{'}}^2}{g_{11}T^2}=c_w\frac{{f{'}^2(r)}g(r)}{4T_0^2f(r)}~;
\nonumber
\\
&& -\frac{c_g }{g_{11}^2T^2}\Big(2\sigma{''} g_{11}-{\sigma}{'} g_{11}{'}\Big)
 = -c_g{T_0}^{-2}\Big[f{''}g-\frac{g{f{'}}^2}{f}+\frac{g{'}f{'}}{2}\Big]~.
\end{eqnarray} 
Next, to take the horizon limit, expand the metric coefficients around the horizon $r=r_0$ in the Taylor series: 
\begin{eqnarray}
f(r) = f'(r_0)(r-r_0) + \dots ; \,\,\,\,\ g(r) = g'(r_0)(r-r_0)+\dots
\label{nearhorioncoff}
\end{eqnarray}
Substituting these in (\ref{barc23}) and then taking the limit we obtain
\begin{eqnarray}
\label{frprim}
&&\lim_{r\rightarrow r_0} c_w\frac{{\sigma{'}}^2}{g_{11}T^2} = c_w\frac{f'(r_0 )g'(r_0 )}{4T_0^2}=c_w\frac{\kappa^2}{T_0^2}=4\pi^2 c_w~;
\nonumber
\\
&&\lim_{r\rightarrow r_0} -\frac{c_g }{g_{11}^2T^2}\Big(2\sigma{''} g_{11}-{\sigma}{'} g_{11}{'}\Big) = c_g{T_0}^{-2}\frac{g{'}(r_0)f{'}(r_0)}{2} = 2c_g\frac{\kappa^2}{T_0^2} = 8\pi^2 c_g~.
\end{eqnarray}
In the above, first the expression for the surface gravity $\kappa = \sqrt{f'(r_0 )g'(r_0 )}/2$ and finally the equilibrium temperature in terms of surface gravity ($T_0 = \kappa/2\pi$) have been used. Therefore (\ref{divh}) becomes 
\begin{eqnarray}
\bar{C}_1+\bar{C_2}\leq (4\pi^2c_w + 8\pi^2c_g)~.
\label{main}
\end{eqnarray}
The above relation is our main result. Such a relation implies that the sum of the response parameters, appearing in the zeroth order in the constitutive relation (\ref{chi2}) in two dimensions, in presence of anomalies, cannot be greater than a linear combination of the anomaly coefficients, appearing in the second order in (\ref{chi2}). Since this condition comes from a very fundamental physical law of thermodynamics, it can be treated as a basic constraint of the present theory.
It is interesting to note that in the reversible situation (when equality holds), the above leads to $\bar{C}_1+\bar{C_2}=(4\pi^2c_w + 8\pi^2c_g)$ which reproduces earlier findings in two ways: one by studying the field theory on a cone \cite{Jensen:2012kj} and other by imposing the {\it Israel-Hartle-Hawking} boundary condition \cite{Bibhas:2014rp,Dey:2014}, both of which yield $\bar{C}_1=4\pi^2c_w$ and $\bar{C_2}=8\pi^2c_g$. It indicates that boundary condition; i.e. the {\it Israel-Hartle-Hawking} state, might be related to the second law of thermodynamics or the vacuum has an important role in the thermodynamics of anomalous fluid. 
\section{\label{sec5}Summary and conclusions}
 
In this paper we have given an exact expression (\ref{TSa}) for the entropy current in $(1+1)$ dimensional hydrodynamics with gravitational anomalies. Further, demanding compatibility with the local form of the second law of thermodynamics, this expression was instrumental in providing a bound (\ref{main}) on the sum of the response parameters in terms of the coefficients of the conformal and diffeomorphism anomalies. In the special case where the bound gets saturated, results obtained in the literature by either analyzing field theory on a cone \cite{Jensen:2012kj} or exploiting the {\it{Israel-Hartle-Hawking}} vacuum \cite{Bibhas:2014rp,Dey:2014} as a boundary condition were reproduced.

In the absence of approaches based on {\it{Landau-Lifshitz}} theory \cite{Landau} or Eckart theory \cite{Eckart:1940te}, which are geared for first order formulations, one has to find alternative ways to find the entropy. Also, Israel-Stewart's \cite{Israel} approach fails since this is applicable for an anomaly free theory. The popular derivative expansion method is of course viable, but as already elaborated, a theory in $(1+1)$ dimensions is rather special since it yields exact solutions. Indeed, following this philosophy, we have developed \cite{Banerjee:2013qha,Banerjee:2013fqa,Bibhas:2014rp,Dey:2014} a systematic scheme to obtain exact constitutive relations for fluids in the presence of both gauge and gravitational anomalies. The results were used here to obtain the cherished form of the entropy current. Indeed, the precise structure of this current (\ref{TSa}) was inspired by the prescription that is followed for the ideal fluid and then completing it by the exact expression for the constitutive relation (\ref{chi2}).

It is interesting to note that when the bound (\ref{main}) gets saturated, the result found by using the {\it Israel-Hartle-Hawking} vacuum as a boundary condition is reproduced. It may be recalled that this particular vacuum choice implies that all out-going and in-going modes of the stress tensor in Kruskal coordinates are regular near the horizon. Translated in thermodynamic language this would imply a reversible process. This is manifested in the bound since, when it gets saturated, a reversible process is implied. In this way our analysis provides a connection between boundary condition and the nature (reversible/irreversible) of a thermodynamic process.    
  
   As a final remark we note that this analysis is mostly general and can be extended for gauge anomaly. We leave this exercise for the future.
\vskip 9mm
\noindent
{\bf Acknowledgments}\\
One of the authors (S.D) thanks S.N. Bose National Centre For Basic Sciences, India for providing Senior research fellowship. The research of B.R.M is supported by Lady Davis and Golda Meir fellowships at Hebrew University, by the I-CORE Program of the Planning and Budgeting Committee and the Israel Science Foundation (Grant No. 1937/12), as well as by the Israel Science Foundation personal Grant No. 24/12. He is also supported by a START-UP RESEARCH GRANT (No.  SG/PHY/P/BRM/01) from Indian Institute of Technology Guwahati, India.

 \end{document}